\documentclass[twocolumn,showpacs,superscriptaddress,preprintnumbers,floatfix,amsmath,amssymb,prb]{revtex4-1}

\usepackage{graphicx,color}% Include figure files
\usepackage{dcolumn}% Align table columns on decimal point
\usepackage{bm}% bold math
\usepackage{amsfonts}% bold math
\usepackage[below]{placeins}
\usepackage{subfigure}
\begin{document}

\title{The magnetic order of a manganese vanadate system with two-dimensional striped triangular lattice.}

\author{V. Ovidiu Garlea}
 \email{garleao@ornl.gov}
\affiliation{Neutron Scattering Division, Oak Ridge National Laboratory, Oak Ridge, Tennessee 37831, USA}
\author{Michael A. McGuire}
\affiliation{Materials Science and Technology Division, Oak Ridge National Laboratory, Oak Ridge, Tennessee 37831, USA}
\author{Liurukara D. Sanjeewa}
\affiliation{Department of Chemistry and Center for Optical Materials Science and Engineering Technologies (COMSET), Clemson University, Clemson, South Carolina 29634-0973, USA}
\affiliation{Materials Science and Technology Division, Oak Ridge National Laboratory, Oak Ridge, Tennessee 37831, USA}
\author{Daniel M. Pajerowski}
\author{Feng Ye}
\affiliation{Neutron Scattering Division, Oak Ridge National Laboratory, Oak Ridge, Tennessee 37831, USA}
\author{Joseph W. Kolis}
\affiliation{Department of Chemistry and Center for Optical Materials Science and Engineering Technologies (COMSET), Clemson University, Clemson, South Carolina 29634-0973, USA}


\begin{abstract}
Results of magnetization and neutron diffraction measurements of the manganese vanadate system Mn$_5$(VO$_4$)$_2$(OH)$_4$ are reported. The crystal structure of this compound contains triangular [Mn$_3$O$_{13}$] building blocks that produce two-dimensional Mn$^{2+}$ magnetic networks with striped triangular topologies. The Mn sheets are connected through the nonmagnetic vanadate tetrahedra extending along the $a$-axis. Magnetization measurements performed on single crystals reveal the onset of a long-range antiferromagnetic order below approximately 45 K. The magnetic structure is N\'{e}el-type with nearest-neighbor Mn atoms coupled via three or four antiferromagnetic bonds. The magnetic moments are confined within the layers and are oriented parallel to the $b$ direction. The magnitudes of ordered moments are reduced, presumably by geometrical frustration and the low-dimensionality of the lattice structure.
\end{abstract}

%\pacs{75.50.Gg, 75.60.Jk, 75.25.-j, 71.27.+a, 61.05.F-}% PACS, the Physics and Astronomy Classification Scheme.

\maketitle

\section{Introduction}

The roles played by frustration and quantum fluctuations on the magnetic ordering of antiferromagnetic systems with spins residing on triangular lattices continue to be the subject of intense research. The most commonly encountered triangular lattices are based on planes of edge-sharing equilateral triangles where each magnetic atom has coordination number $z$ = 6, and competing nearest-neighbor (NN) interactions destabilize the magnetic order.\cite{Collins,Starykh} A lower $z$ = 3 coordination number is found in the layered honeycomb lattices, which are derived from the edge-sharing triangular planar lattice by removing one third of the magnetic sites in an ordered fashion. In the layered honeycomb case, the frustration arises from competing interactions rather than geometric constraints. Depending on the relative strengths of first, second and third neighbor magnetic exchange interactions, several antiferromagnetic-ordering patterns are possible, including N\'{e}el, zigzag, stripy, and spiral orders.\cite{honeycomb1,honeycomb2} An increase in complexity of the magnetic phase diagram is expected for 1/6 depleted triangular lattices, where additional magnetic ions fill the center of some of hexagonal honeycombs to produce striped triangular patterns characterized by several possible coordination numbers $z$ = 4, 5 and 6. This type of magnetic lattice was found in the Cu$_5$(VO$_4$)$_2$(OH)$_4$ (turanite) layered compound, exhibiting edge-sharing Cu(II)O$_6$ octahedra that form alternate packing of triangular and honeycomb strips. Recent magnetic measurements have shown that this material behaves as a spin-1/2 ferrimagnet and exhibits an unusual 1/5 magnetization plateau arising from the competition between antiferromagnetic and ferromagnetic interactions caused by the strong frustration.\cite{turanite} Similar magnetic lattice has recently been found in the manganese vanadate congener Mn$_5$(VO$_4$)$_2$(OH)$_4$.~\cite{duminda} We report here a comprehensive investigation of anisotropic magnetic properties and magnetic structure of the manganese material by means of magnetization and neutron diffraction measurements. We show that this system orders in an antiferromagnetic state with non-uniform static moment distribution and it exhibits a field-induced spin-reorientation transition.

\begin{figure}[tbp]
\includegraphics[width=3.4in]{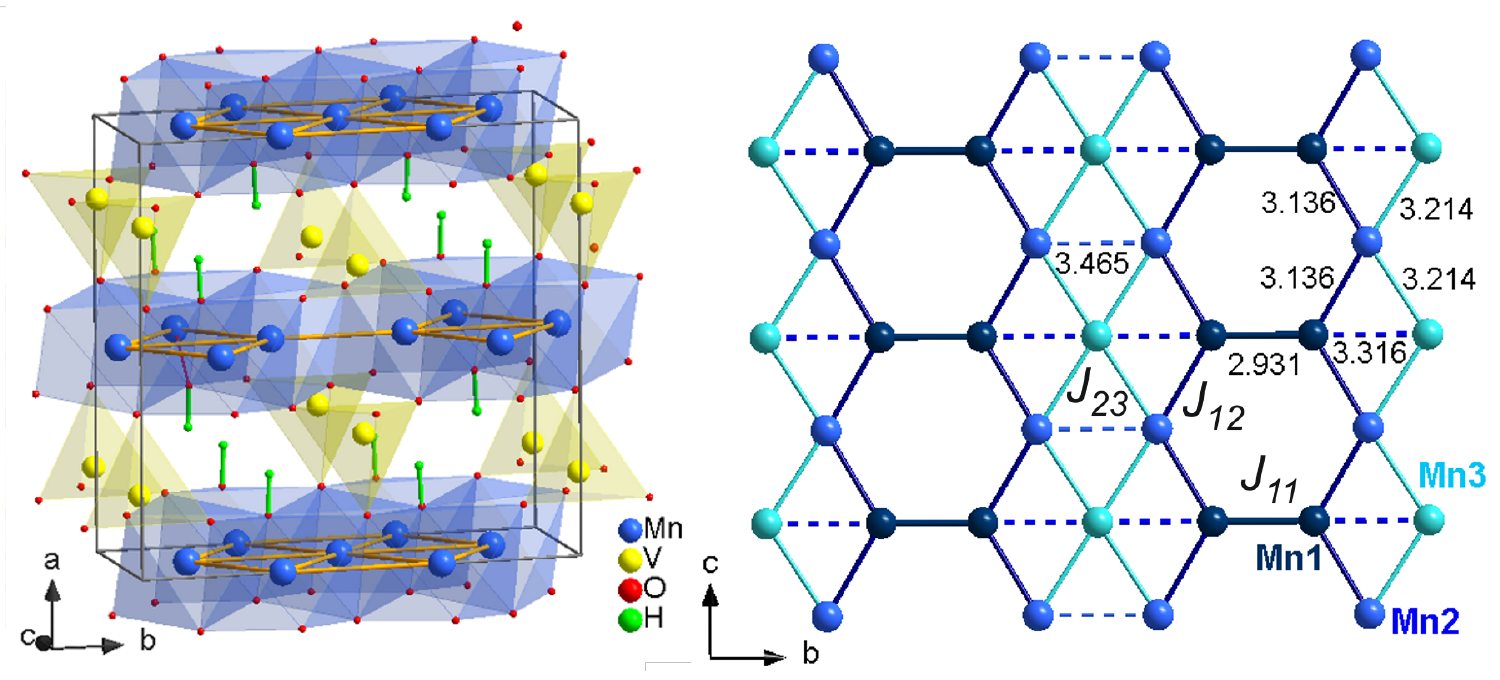}
\caption{\label{structure} (a) Polyhedral view of the crystal structure of Mn$_5$(VO$_4$)$_2$(OH)$_4$. (b) The striped triangular magnetic lattice formed of three crystallographically distinct Mn ions in the $bc$ plane.}
\end{figure}

\section{Experimental details}

Single crystals of Mn$_5$(VO$_4$)$_2$(OH)$_4$ were grown using a high-temperature hydrothermal technique in sealed silver ampules, as described in Ref.~\onlinecite{duminda}. For the present study we used a single crystal with a rectangular shape of approximately 1.5 x 2 x 5 mm$^3$.
Temperature and field-dependent magnetic measurements were carried out with a Quantum Design Magnetic Property Measurement System (MPMS). The measurements were taken in the magnetic fields up to 50 kOe, applied parallel to the $a$ and $b$ crystallographic directions, at temperatures from 2 K to 300 K. Neutron diffraction measurements were performed at the CORELLI\cite{corelli} instrument at the Spallation Neutron Source. White-beam Laue diffraction measurements were taken at four different temperatures 5~K, 20K, 70 K and 200 K. For each temperature, the sample was rotated in steps of 3$^\circ$ over a ranges of 180$^\circ$. Data correction and reduction was carried out using MANTID software.\cite{mantid} Magnetic structures models have been constructed by representation analysis using the program SARA{\it h},\cite{sarah} and structural and magnetic data refinements were performed with the FullProf Suite program.~\cite{fullprof}

\section{Results and Discussion}
\subsection{Crystal structure}

\begin{table}[tbp]
\caption{\label{table1} Nearest neighbor distances between magnetic ions in Mn$_5$(VO$_4$)$_2$(OH)$_4$, as reported in Ref.~\onlinecite{duminda}}
\begin{ruledtabular}
\begin{tabular}{r p{1.0in} c}
Mn1 - Mn1 (x1) :& 2.931(1)~\AA~~$\rightarrow$ $J_{11}$\\[3pt]
    - Mn2 (x2) :& 3.136(5)~\AA~~$\rightarrow$ $J_{12}$\\[3pt]
    - Mn3 (x1) :& 3.316(1)~\AA~~$\rightarrow$ $J_{13}$\\[3pt]
Mn2 - Mn1 (x2) :& 3.136(5)~\AA~~$\rightarrow$ $J_{12}$\\[3pt]
    - Mn2 (x1) :& 3.464(1)~\AA~~$\rightarrow$ $J_{22}$\\[3pt]
    - Mn3 (x2) :& 3.213(1)~\AA~~$\rightarrow$ $J_{23}$\\[3pt]
Mn3 - Mn1 (x2) :& 3.316(1)~\AA~~$\rightarrow$ $J_{13}$\\[3pt]
    - Mn2 (x4) :& 3.213(1)~\AA~~$\rightarrow$ $J_{23}$\\[3pt]
\end{tabular}
\end{ruledtabular}
\end{table}

A polyhedral view of the Mn$_5$(VO$_4$)$_2$(OH)$_4$ crystal structure along with a view of an isolated Mn layer are shown in Fig.~\ref{structure}. Mn$_5$(VO$_4$)$_2$(OH)$_4$ crystallizes in a monoclinic unit cell of symmetry C$2/m$ and dimensions $a$ = 9.6568(9)~\AA, $b$ = 9.5627(9)~\AA, $c$ = 5.4139(6)~\AA, $\beta$ = 98.529(8)$^\circ$ at T = 300 K.~\cite{duminda} The structure consists of two-dimensional layers in the $bc$ plane of edge-sharing distorted octahedral [MnO$_4$(OH)$_2$] and [MnO$_2$(OH)$_4$] units. In each layer, there are three crystallographically distinct Mn$^{2+}$ sites: Mn1 (Wyckoff site: 4h), Mn2 (4g) and Mn3 (2d). The Mn1 and Mn2 sites have similar octahedral coordinations ([MnO$_4$(OH)$_2$]) and form a honeycomb lattice, whereas the Mn3 atoms (in [MnO$_2$(OH)$_4$] coordination) fill the center of every second honeycomb along the $b$ direction to create a striped triangular lattice. In this arrangement, each Mn site has a different NN coordination number. As indicated in Table~\ref{table1}, Mn1 has four NNs, Mn2 has five NNs, and Mn3 is surrounded by six NNs. Based on these interatomic bonds one can define NN intra-layer exchange interactions $J_{ij}$ ($i$,$j$=1,2,3). The Mn layers are connected through the vanadate tetrahedra (VO$_4$) extending along the $a$-axis. Due to the non-magnetic nature of V$^{5+}$ cation, the magnetic interactions between adjacent Mn layers ($J_z$), distanced by $a$/2 $\simeq$ 4.8~\AA, are expected to be weaker compared to the interactions within the layers. No significant change in crystal structure was observed during sample cooling from the room temperature to 5 K.

\subsection{Macroscopic magnetic properties}

\begin{figure}[tbp]
\includegraphics[width=3.4in]{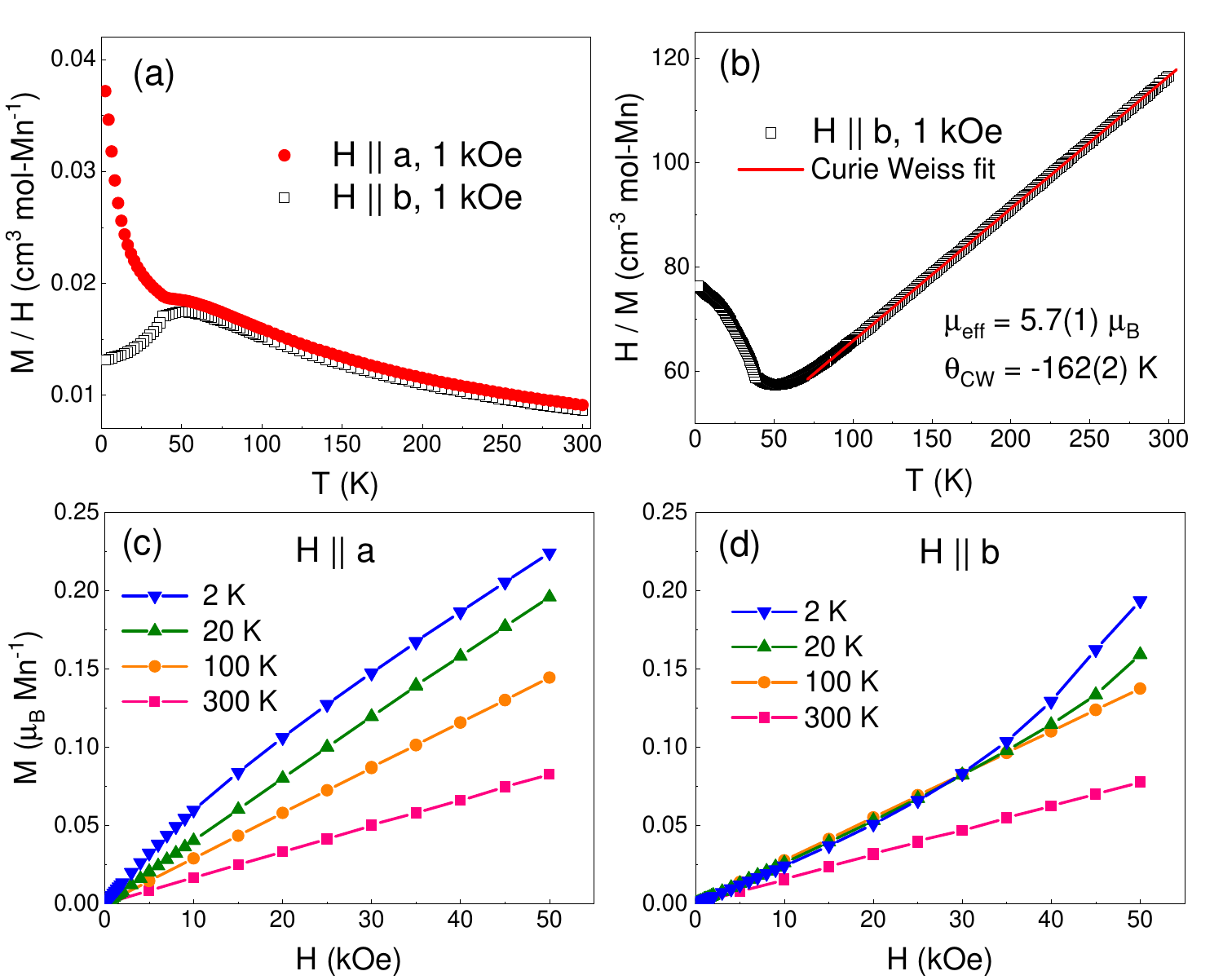}
\caption{\label{magnetization} (a) Evolution of the magnetic susceptibility ($M/H$) as a function of temperature, for a field of 1 kOe applied along $a$ (solid red symbol) and $b$ -axis (open square). (b) Plot of the inverse susceptibility as a function of temperature and the linear Curie-Weiss fit. (c) and (d) Isothermal magnetization curves measured at selected temperatures ranging from 2 K to 300 K for magnetic fields applied parallel to $a$ and $b$ -axes, respectively. A  spin-reorintation transition is observed near 35 kOe, when the field is applied along $b$ -axis.}
\end{figure}

The temperature dependence of static magnetic susceptibility ($\chi=M/H$) measured with an applied field of 1 kOe along $a$ and $b$ directions is shown in Fig.~\ref{magnetization}(a). The susceptibility curve exhibits for $H \parallel b$ a peak at about 45 K, indicating antiferromagnetic long-range ordering. The strong divergence between the curves measured with magnetic fields applied along the two crystallographic directions suggests that the ordered magnetic moments are predominantly along the $b$ direction. The low-temperature divergence of susceptibility along the $a$-axis may be due to a small canting of the moments along that direction. Our single crystal data compares nicely with the powder data taken with H = 10 kOe and shown in Ref~\onlinecite{duminda}. The inverse susceptibility (1/$\chi$), displayed in Fig.~\ref{magnetization}(b), follows Curie-Weiss behavior above approximately 100 K. The fit using the Curie-Weiss model yields a Weiss temperature of -162(2) K, due to dominant antiferromagnetic interactions between the Mn moments. The determined effective magnetic moment of 5.7(1)~$\mu_B$/Mn$^{2+}$ is consistent with the expected value of $g \sqrt{S(S+1)}$ = 5.91~$\mu_B$/Mn for high spin $S$ =5/2. The deviation of 1/$\chi$ below 100 K, indicates the existence of magnetic correlations well above the ordering temperature. The ratio $f=|\Theta_{CW}|/T_N$ = 3.6 suggests that magnetic frustration is present in this material.

The isothermal magnetization curves at 2 K, 20 K, 100 K and 300 K in magnetic fields up to 50 kOe applied along $a$ and $b$ directions are shown in Figs.~\ref{magnetization}(c) and (d). When the field is applied along the $a$ axis the magnetization shows a smooth increase with increasing magnetic field. In contrast, for an applied field along the $b$ axis, the magnetization curves measured below the T$_N$ ordering temperature exhibit a step-like transition near approximately 35 kOe. This is likely associated with a spin-reorientation transition, with moments rotating towards the $ac$ plane. A similar spin-flop transition has been seen to occur in honeycomb magnetic lattice ordered in a N\'{e}el-type AFM ground state.\cite{Li2MnO3}

\subsection{Magnetic structure}

\begin{figure}[tbp]
\includegraphics[width=3.45in]{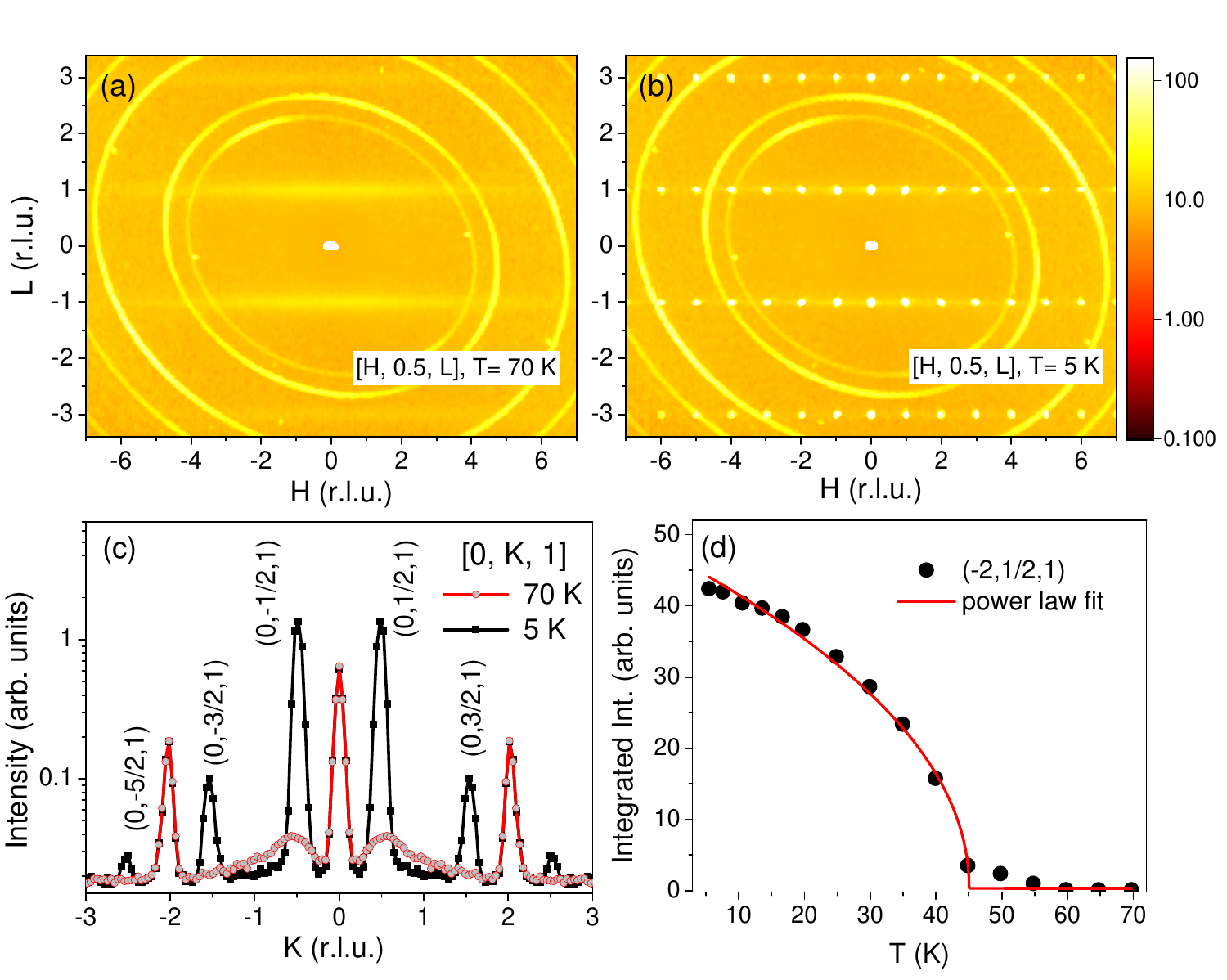}
\caption{\label{diffraction} (a) Data collected at 70 K, using Corelli instrument, that shows rods of diffuse scattering along the $H$-direction at $K$ = 1/2 and $L$ = $\pm$1, $\pm$3 positions. (b) Data collected at 5 K showing magnetic Bragg peaks corresponding to a propagation vector $k$=(0,~1/2,~0). The stripes of diffuse scattering along $H$ remain visible. (c) Comparison of [0, $K$ ,1] - cuts from 70 K and 5 K data. (d) Temperature dependence of (-2, 1/2, 1) magnetic peak intensity and the power-law fit.}
\end{figure}

Single crystal neutron diffraction data collected at 200 K, in the paramagnetic regime, is well accounted by the structural model introduced by Sanjeewa.\cite{duminda} Data collected at 70 K, shown in Fig.~\ref{diffraction}(a), reveals a pattern of diffuse scattering taking the form of rods of scattering along the $H$-direction at $K$ = 2$n$+1/2 ($n$ = integer) and $L$ = odd positions. The scattering intensity along $Q_H$ follows the decay expected for the Mn$^{2+}$ magnetic form factor. This diffuse scattering indicates the formation of strong two-dimensional magnetic correlations in Mn layers ($bc$-plane) well above the long-range ordering temperature. New satellite magnetic Bragg peaks corresponding to a propagation vector $k$=(0,~1/2,~0) appear in the data measured at 5~K, as displayed in Fig.~\ref{diffraction}(b). The stripes of diffuse scattering along $H$ remain visible, although with reduced intensity. This persistence of diffuse scattering suggests that a certain stacking disorder of the magnetic moments continue to exist even in the long-range ordered state. A comparison of the $Q = [0, K ,1]$ - cut of 70 K and 5 K data is shown in Fig.~\ref{diffraction}(c), where the resolution limited magnetic peaks at $K$ = 2$n$+1/2 are visible. The evolution with temperature of the (-2, 1/2, 1) magnetic peak intensity is presented in Fig.~\ref{diffraction}(d). The magnetic order parameter can be described by a power-law curve $I(T)\propto (1-T/T_N)^{2\beta}$ with $T_N$ = 45 K and  a critical exponent $\beta$ = 0.25(1). This value is close to the exponent 0.23 that was shown to be a universal signature of finite-sized 2D XY behaviour.\cite{bramwell} The extension of the goodness of fit beyond the critical regime is typical for low dimensional magnets.

Symmetry-allowed magnetic structures that can result from a second-order magnetic phase transition, given the crystal structure C$2/m$ space group and $k$=(0,~1/2,~0) were examined using the program SARA{\it h}.\cite{sarah} The calculations showed that there are two possible irreducible representations (IRs) that constrain the magnetic moments to be either parallel to $b$ - axis or contained in the $ac$ plane.  The Mn1 and Mn2 atoms positions ($4h$ and $4g$) are split into two orbits as there are insufficient symmetry elements that leave the propagation vector ${\mathbf k}$ invariant to generate all equivalent positions. Nonetheless, Mn moments of crystallographically similar sites were constrained to have the same magnitudes. We found that the collinear structural model with moments aligned parallel to the $b$ -axis gives the best fit to the neutron diffraction intensities and also explains the magnetization results.

\begin{figure}[tbp]
\includegraphics[width=2.7in]{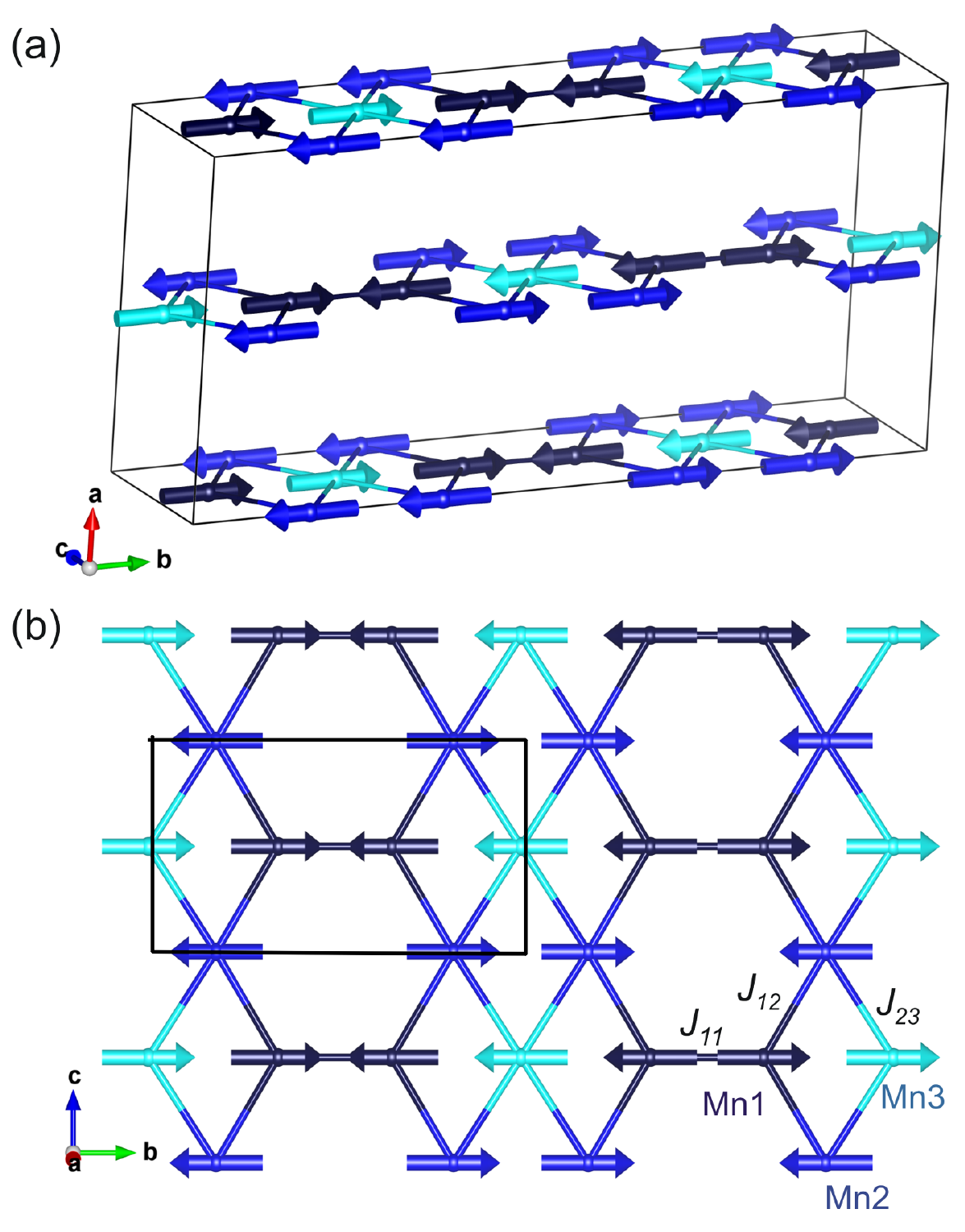}
\caption{\label{magstr} Stereographic view of the magnetic structure (a) and the magnetic configuration inside the $bc$-plane (b). The Mn1 ordered moments are stabilized by three NN AFM interactions ($J_{11}$ and two $J_{12}$), while the Mn2 and Mn3 moments are stabilized via four NN AFM interactions (two $J_{12}$ and two $J_{23}$}
\end{figure}

In the selected magnetic structure model the magnetic moments at Mn2 positions are oriented antiparallel with respect to those at Mn1 and Mn3 positions. The Mn1 and Mn2 moments distributed on a honeycomb morphology exhibit a collinear N\'{e}el order with antiferromagnetic nearest-neighbor interactions $J_{11}$ and $J_{12}$. Inside the triangular strips, groups of ferromagnetic Mn1 and Mn3 moments couple antiferromagnetically with nearest-neighbor Mn2 moments along the $c$-direction via $J_{12}$ and $J_{23}$ exchange interactions. The magnetic ordering configuration inside the chemical unit-cell, propagating with the wave-vector $k$=(0,~1/2,~0), is explicitly given in Table~\ref{table2}. This spin configuration can be described by the magnetic space group P$_b2/c$ in a magnetic cell that is doubled along the $b$-axis. A stereographic view of the magnetic structure and the in-layer ($bc$-plane) magnetic configuration are shown in Fig.~\ref{magstr}. As visible in the figure, the Mn1 ordered moments are stabilized by three NN AFM bonds, while Mn2 moments are stabilized via four NN AFM bonds. The Mn3 is surrender by two Mn1 and four Mn2 atoms and its ordered moment is expected to be affected by those interactions. The refined magnitudes of the ordered moments obtained from 183 magnetic Bragg intensities are: 1.98 $\pm$ 0.07 $\mu_B$ for Mn1, 2.43 $\pm$ 0.05 $\mu_B$ for Mn2 and 2.48 $\pm$ 0.09 $\mu_B$ for Mn3. These values are significantly lower than those expected for the Mn$^{2+}$ ions with $S$=5/2. The non-uniformity of the ordered magnetic moment can be due to the different degree of magnetic frustration for the three distinct positions. It appears that the different ordered moment is not correlated to the dissimilarity in the octahedral coordination of the three Mn sites, but rather to the number of NN AFM bonds. In addition, the reduced moments magnitude can also be attributed to the strong two-dimensionality of the magnetic structure, evidenced by coexistence of Bragg reflections with the diffuse scattering, that may arise from frustrated inter-plane interactions.

\begin{table}[tbp]
\caption{\label{table2} Magnetic moment arrangement in the AFM state of Mn$_5$(VO$_4$)$_2$(OH)$_4$. The atomic positions correspond to the chemical lattice, and the associated moments propagate with the wave-vector $\textbf{k}$=(0, 1/2, 0). This spin configuration is described by the magnetic space group P$_b2/c$ in a doubled magnetic cell along the $b$-axis.}
\begin{ruledtabular}
\begin{tabular}{ccc}
\vspace{.1in}
Atom & ( $x$, $y$, $z$ )&( $m_a$, $m_b$, $m_c$ )\\[3pt]
\hline
Mn1$_1$ &( 0,~0.153,~0.5 )&( 0,~$m_1$,~0 )\\[3pt]
Mn1$_2$ &( 0,~0.846,~0.5 )&( 0,~$m_1$,~0 )\\[3pt]
Mn2$_1$ &( 0,~0.681,~0.0 )&( 0,~$-m_2$,~0 )\\[3pt]
Mn2$_2$ &( 0,~0.318,~0.0 )&( 0,~$-m_2$,~0 )\\[3pt]
Mn3 &( 0,~0.5,~0.5 )&( 0,~$m_3$,~0 )\\[3pt]
\end{tabular}
\end{ruledtabular}
\end{table}

\section{Summary}

The magnetic properties of the manganese vanadate Mn$_5$(VO$_4$)$_2$(OH)$_4$ have been studied by means of magnetization and neutron powder diffraction measurements. The crystal structure of this system is composed of two-dimensional layers where Mn$^{2+}$ atoms form honeycomb-like hexagons. Every other row of hexagons along the $b$ direction contains an additional Mn atom in the center to produce a striped triangular lattice. Magnetization measurements indicate a magnetic order at 45 K and a field-induced spin-reorientation transition at approximately 35~kOe for a magnetic field applied along the $b$-direction. Neutron diffraction experiments performed in zero magnetic field confirm the antiferromagnetic magnetic order below 45~K, and evidence the coexistence of the long-range ordered and two-dimensional (long-range disordered) states. The ordered state consists of a collinear N\'{e}el -type arrangement of Mn moments oriented parallel to the $b$ direction. The three distinct Mn sites contained in each magnetic layer are found to order with different moment magnitudes, in consent with the number of their nearest-neighbor AFM couplings. The strongly reduced ordered magnetic moments extracted from the data refinement demonstrate the strong two-dimensionality of the system and the existence of a magnetically frustrated ground state.

\begin{acknowledgments}
Work at the Oak Ridge National Laboratory, was sponsored by the US Department of Energy, Office of Basic Energy Sciences, Materials Sciences and Engineering Division (for macroscopic characterization) and Scientific User Facilities Division (for neutron scattering studies). The authors also acknowledge the financial support from the National Science Foundation under grant no. DMR-1410727. Authors thanks A. Savici for the help provided with the CORELLI data processing.

Notice: This manuscript has been authored by UT-Battelle, LLC under Contract No. DE-AC05-00OR22725 with the U.S. Department of Energy. The United States Government retains and the publisher, by accepting the article for publication, acknowledges that the United States Government retains a non-exclusive, paid-up, irrevocable, world-wide license to publish or reproduce the published form of this manuscript, or allow others to do so, for United States Government purposes. The Department of Energy will provide public access to these results of federally sponsored research in accordance with the DOE Public Access Plan (http://energy.gov/downloads/doe-public-access-plan).
\end{acknowledgments}

\end{document}